# Impact of Surface Roughness in Measuring Optoelectronic Characteristics of Thin-Film Solar Cells

David Magginetti[1], Seokmin Jeon[2], Yohan Yoon[2,3], Ashif Choudhury[1], Ashraful Mamun[1], Yang Qian[1], Jordan Gerton[4], and Heayoung Yoon[1]

[1] Department of Electrical and Computer Engineering, University of Utah, UT 84112, USA
[2] US Naval Research Laboratory, Washington, DC 20375, USA
[3] Department of Materials Engineering, Korea Aerospace University, Goyang, 10540 South Korea
[4] Department of Physics and Astronomy, University of Utah, UT 84112, USA

*Abstract* — Microstructural properties of thin-film absorber layers play a vital role in developing high-performance solar cells. Scanning probe microscopy is frequently used for measuring spatially inhomogeneous properties of thin-film solar cells. While powerful, the nanoscale probe can be sensitive to the roughness of samples, introducing convoluted signals and unintended artifacts into the measurement. Here, we apply a glancing-angle focused ion beam (FIB) technique to reduce the surface roughness of CdTe while preserving the subsurface optoelectronic properties of the solar cells. We compare the nanoscale optoelectronic properties "before" and "after" the FIB polishing. Simultaneously collected Kelvin-probe force microscopy (KPFM) and atomic force microscopy (AFM) images show that the contact potential difference (CPD) of CdTe pristine (peak-to-valley roughness > 600 nm) follows the topography. In contrast, the CPD map of polished CdTe (< 20 nm) is independent of the surface roughness. We demonstrate the smooth CdTe surface also enables high-resolution photoluminescence (PL) imaging at a resolution much smaller than individual grains (< 1 µm). Our finite-difference time-domain (FDTD) simulations illustrate how the local light excitation interacts with CdTe surfaces. Our work supports low-angle FIB polishing can be beneficial in studying buried sub-microstructural properties of thin-film solar cells with care for possible ion-beam damage near the surface.

## I. Introduction

Thin-film CdTe solar cells are a leading photovoltaic (PV) technology owing to cost-effective manufacturing and reliable power production for 20+ years [1, 2]. Optimized close-space sublimation (CSS) and vertical traveling deposition (VTD) enable the rapid production of high-quality CdTe thin films. Such absorber layers consist of inhomogeneous microstructures with a typical grain size of approximately 1 µm after post-deposition processing. Previous studies suggested that electrically-active point defects and structural defects of microstructures (i.e., grain bulk, grain boundaries) can significantly impact the cell efficiencies and long-term stability and reliability of CdTe solar cells [3, 4].

Scanning probe microscopy (SPM) has been extensively used for characterizing inhomogeneous semiconductor materials and devices. Examples include atomic force microscopy, electron-beam microscopy, and confocal optical microscopy, where a local excitation source is raster-scanned on the area of interest to image spatially resolved properties of solar cells [5-9]. While powerful, the nanoscale probe used in SPM can be sensitive to the surface roughness of the sample, introducing convoluted signals and unintended artifacts into the measurement. Various polishing techniques have been proposed and applied to reduce the sample roughness, including focused ion beam (FIB), gas-assisted etching, wet etching, and low-temperature milling [10-12]. Among them, previous efforts demonstrated an optimized argon ion ($Ar^+$) beam could efficiently produce a smooth surface with minimizing possible beam damage [13, 14].

In this work, we use glancing-angle FIB milling to reduce the topographical variation of CdTe while retaining the quality of the sample. AFM/KPFM is used to measure the surface potential of CdTe "before" and "after" the polishing. Analysis of KPFM shows the different CPD profiles of pristine and polished devices. We demonstrate PL imaging that resolves the luminescence characteristics of individual grains and grain boundaries. We perform FDTD simulations to visualize the light absorption profile of CdTe microstructures, illustrating the non-uniform interactions of the local light source to grains and grain boundaries in nanoscale PL imaging.

## II. Experimental

This study used conventional CdS/CdTe solar cells extracted from a solar panel [6]. Figure 1(a) shows a representative topographic image of the pristine CdTe obtained by atomic force microscopy (AFM). Irregular grain sizes range from < 1 µm to a few µm with a peak-to-valley variation as high as 0.6 µm in this sampling area (10 µm × 10 µm). To reduce the roughness, we performed $Ar^+$ polishing at an incident beam energy of 3 keV for one hour. The ion beam was irradiated at a glancing angle of 1°, only removing a thin layer of CdTe while preserving the CdTe bulk unchanged (Fischione 1060). A schematic in Figure 1(b) illustrates the ion-beam milling process. KPFM (CPD) and AFM (topography) images were simultaneously collected on the samples of "before (pristine)" and "after (polished)" the milling (Bruker AFM probes; OSCM-PT-R3). The diameter of a cantilever tip was 20 nm, and the resonance frequency of the tip was set to ≈ 71 kHz with a spring constant of 2 N/m. PL imaging at 405 nm was performed using a customized confocal microscopy system. The PL maps were collected via immersion objective lens (100x) in an oil media (Cargille type B). Finite-difference time-domain (FDTD) simulations were conducted to study the absorption profiles of the laser beam in CdTe for both smooth and rough sample surfaces (Ansys Lumerical). FDTD methods are often used to solve electromagnetic interaction in a sample of interest. Maxwell equations are solved numerically on a discrete grid (mesh) in space and time defined in the model. We used a standard material library (J. A. Wollam Ellipsometry Solutions) to formulate the refractive index and extinction coefficient of CdTe at different wavelengths. The mesh size was set to 0.25 nm for the simulations.

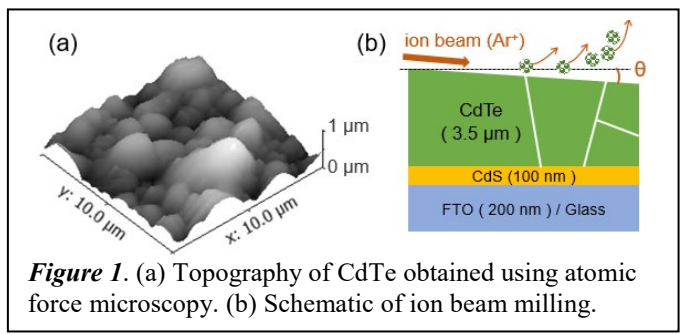

*Figure 1*. (a) Topography of CdTe obtained using atomic force microscopy. (b) Schematic of ion beam milling.

## III. RESULTS AND DISCUSSION

Nanoscale SPM measurements were performed on two CdTe solar cells extracted from the same solar panel. To polish one of the samples, we used a 3 kV $Ar^+$ beam that is irradiated at an angle of 1° parallel to the CdTe surface (Figure 1b). Our milling conditions intend to remove the prominent structures on CdTe (≈ 100 nm) rather than to produce an atomically smooth surface, maintaining a low-level surface roughness after the polishing. Figures 2(a, b) show the topography of the "before (pristine)" and the "after (polished)" CdTe surface. The maximum peak-to-valley roughness of ≈ 600 nm of pristine CdTe was noticeably reduced to ≈ 80 nm after the milling.

The KPFM maps in Figures 2(c, d) show the potential difference of CdTe microstructures. It is apparent that the potential distribution (i.e., CPD) of the pristine CdTe closely follows the topography. Low CPDs are seen on the hills of grains (≈ 1.020 V), while grain boundaries show higher CPDs (≈ 1.035 V) compared to their adjacent grain interiors. We observe the overall CPDs decreases to approximately 30 mV (Figure 2d) after the $Ar^+$ beam milling. This reduction could be attributed to the eliminated $TeO_x$-rich CdTe surface, which exposes the bare CdTe subsurface. Unlike the pristine, the CPD variation of the polished CdTe is independent of the remaining surface roughness, marked with the red boxes in the area in Figure 2. The CPD difference between grain interior and grain boundary of the polished CdTe (≈ 20 mV) is similar to the pristine. The CPD distribution near grain boundaries observed in our KPFM shows a good agreement with the previous work by Jiang *et al.*, where a CdTe sample was polished with a 4 kV $Ar^+$ beam followed by 250 °C annealing [5].

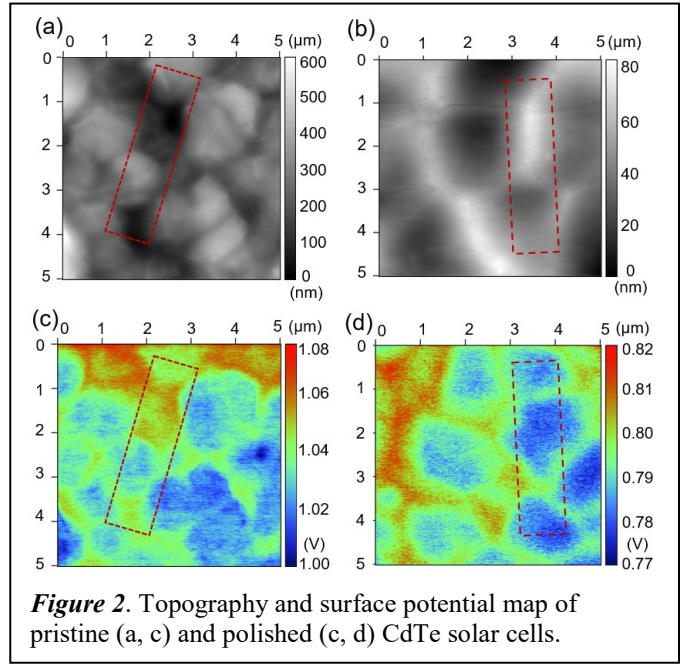

*Figure 2*. Topography and surface potential map of pristine (a, c) and polished (c, d) CdTe solar cells.

The polished CdTe surface enables high-resolution PL imaging with a spatial resolution much smaller than individual grains (< 1 µm). In our confocal PL system, a 100x objective (numerical aperture [NA] = 1.4) focused the collimated laser beam (λ = 405 nm) on CdTe, which was facing down on a cover glass. We used an immersion lens oil (n = 1.5) to match the refractive indices in the PL setup (immersion objective lens / oil / cover glass / CdTe), improving the resolution.

Figure 3 compares the representative PL images obtained on pristine and polished CdTe solar cells. Overall, the PL emission near grain boundaries is lower than grain interiors for both samples. This observation is consistent with previous studies, where the defects near grain boundaries serve as non-radiative recombination centers reducing the PL emission near the bandgap [9, 15]. The brightness contrast of the pristine PL map appears to follow the topography of CdTe microstructures, with random bright spots on the edge of some grains. In contrast, the PL emission of the polished sample shows a relatively uniform contrast between the grain bulk and grain boundary of CdTe. The bright spots are mainly observed in the center of the grain bulk, decaying proportionally to the adjacent grain boundaries. The dark areas in the PL map are likely associated with the concentrated defects rather than unintended artifacts introduced by the topography.

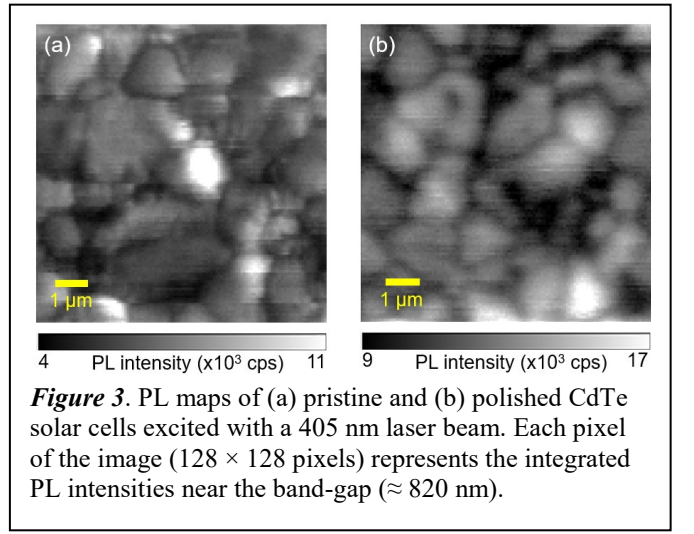

*Figure 3*. PL maps of (a) pristine and (b) polished CdTe solar cells excited with a 405 nm laser beam. Each pixel of the image (128 × 128 pixels) represents the integrated PL intensities near the band-gap (≈ 820 nm).

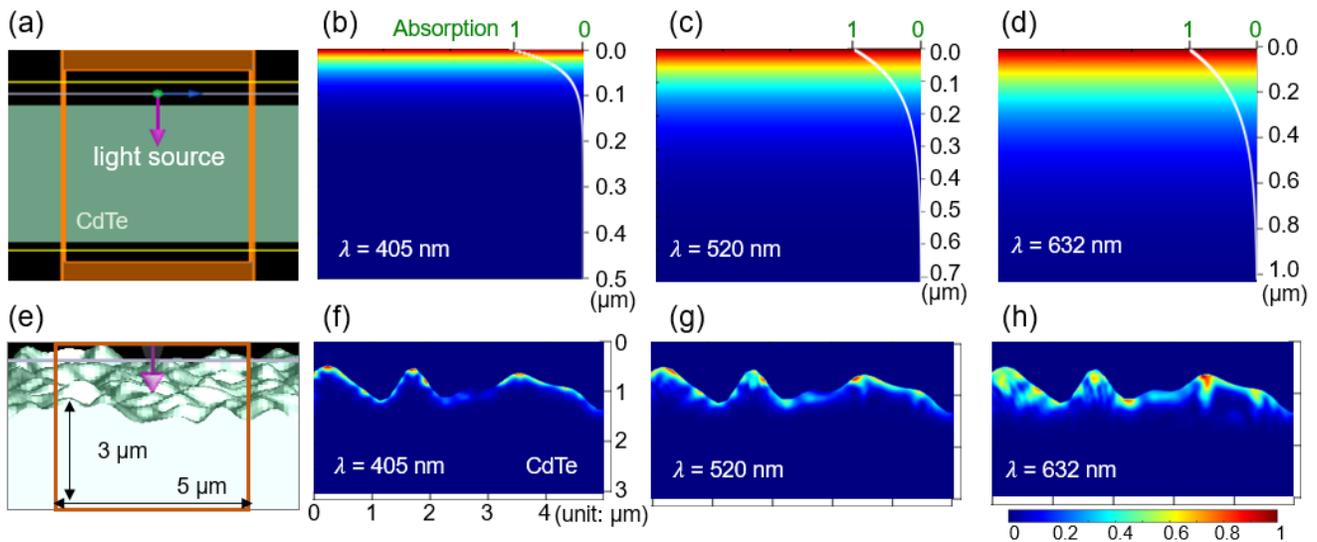

*Figure 4*. Absorption profiles of planar and rough surface CdTe at 405 nm laser beam illuminations. FDTD simulation regions (orange color box in a, e) include the perfect match layer (PMLs) and periodic boundary conditions.

We developed FDTD models of the smooth and rough surface CdTe to understand how the light source (i.e., a laser beam) interacts with the topography, coupled with inhomogeneous microstructures, affects the high-resolution PL imaging. Schematics of the models are shown in Figures 4 (a, e). Our model defined a rough surface CdTe based on the AFM surface roughness seen in Figure 1(a). In addition to the light source of 405 nm used in the PL imaging (Figure 3), we also simulated the absorption profile for 520 nm and 632 nm, frequently used in PL measurements. The absorption profiles of the rough surface are compared to their planar counterparts in Figure 4. As expected, the light absorption decreases exponentially in the planar structures, where the estimated absorption depth of approximately 39 nm, 113 nm, and 215 nm at an illumination of 405 nm, 520 nm, and 632 nm, respectively. Figure 4 (f ~ h) compares the representative absorption landscapes simulated for the rough CdTe surface. It is apparent that the extruded structures of the CdTe film could absorb more photons from an irradiating laser beam. The light absorption near grain boundaries can be influenced by the topography of adjacent grains due to light scattering. Our simulations support that the random bright spots observed in the PL image of the rough CdTe can be attributed to its topography rather than its intrinsic properties.

## IV. CONCLUSIONS

In summary, we have shown the impact of surface roughness in measuring nanoscale optoelectronic characteristics of CdTe solar cells. By reducing the surface roughness using shallow-angle ion beams, the KPFM probe can measure the subsurface CPD distribution of CdTe solar cells. The polished surface enables PL imaging at a spatial resolution much smaller than individual grains (< 1 µm). Our FDTD simulations show inhomogeneous light absorption on rough CdTe, indicating the impact of surface roughness in nanoscale optoelectronic characterizations. Our work supports low-angle FIB polishing can be beneficial in studying buried sub-microstructural properties of thin-film solar cells with care for possible ion-beam damage near the surface.

ACKNOWLEDGEMENT

This research was supported by the U.S. Department of Energy's Office of Energy Efficiency and Renewable Energy (EERE) under the DE-FOA-0002064 program award number DE-EE0008983.We acknowledge support in part by the National Science Foundation (NSF) CAREER Award No. 2048152. This work was support by the USTAR shared facilities at the University of Utah, in part, by the MRSEC Program of NSF under Award No. DMR-1121252. We thank P. Perez, M. Wang, and Z. Liu for valuable inputs and training of the measurement systems.